\documentclass[11pt,onecolumn,draftcls]{IEEEtran}
\newlength{\pic}
\newlength{\bigPic}
\usepackage{graphicx}

\begin{document}

\title{Tandem Coding and Cryptography on Wiretap Channels: EXIT Chart Analysis
\footnote{
Willie K Harrison and Steven W. McLaughlin are with the School of ECE, Georgia Institute of Technology, Atlanta, GA.\\
\indent This work has been accepted to 2009 IEEE International Symposium on Information Theory, June 28-July 3, 2009.}
}
\author{\authorblockN{Willie K Harrison and Steven W. McLaughlin}}

\maketitle

\begin{abstract}
Traditional cryptography assumes an eavesdropper receives an error-free copy of the transmitted ciphertext. Wyner's wiretap channel model recognizes that at the physical layer both the intended receiver and the passive eavesdropper inevitably receive an error-prone version of the transmitted message which must be corrected prior to decryption. This paper considers the implications of using both channel and cryptographic codes under the wiretap channel model in a way that enhances the \emph{information-theoretic} security for the friendly parties by keeping the information transfer to the eavesdropper small. We consider a secret-key cryptographic system with a linear feedback shift register (LFSR)-based keystream generator and observe the mutual information between an LFSR-generated sequence and the received noise-corrupted ciphertext sequence under a known-plaintext scenario. The effectiveness of a noniterative fast correlation attack, which reduces the search time in a brute-force attack, is shown to be correlated with this mutual information. For an iterative fast correlation attack on this cryptographic system, it is shown that an EXIT chart and mutual information are very good predictors of decoding success and failure by a passive eavesdropper.

\end{abstract}

\section{Introduction} Typical communication systems use cryptographic primitives at a layer above the physical layer to achieve data security. Considering that both friendly and eavesdropping parties inevitably receive codewords corrupted by errors at the (noisy) physical layer, Wyner introduced the wiretap channel model \cite{Wyner75} where it was subsequently shown that error control coding (ECC) can contribute to the security against a passive eavesdropper. In this paper we consider the practical situation where both error correction coding and cryptography are used in a wiretap setting and show how the channel errors significantly inhibit the ability of the eavesdropper to recover the cryptographic key using well-known attacks.  We apply the widely used tool of EXIT charts to the wiretap channel and show how they can be used to predict a threshold behavior on the eavesdropper's ability to recover the cryptographic key using a known-plaintext iterative attack.

\begin{figure}
  \begin{center}
  \includegraphics[width=\pic]{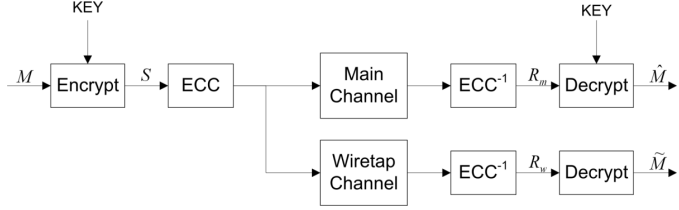}\\
  \end{center}
  \caption{The wiretap channel model shows friendly parties communicating over a main channel, and an eavesdropper observing communications through a wiretap channel.}\label{fig:wiretapChannel}
\end{figure}

The wiretap channel model that employs coding and cryptography is given in Fig. \ref{fig:wiretapChannel}.  In this paper we assume a simple XOR-based cryptographic system whose key is generated using linear feedback shift registers (LFSR). The noisy channel model assumes two discrete memoryless binary symmetric channels (BSC). The \emph{main} channel links friendly parties, while the \emph{wiretap} channel represents a passive eavesdropper observing a noise-corrupted version of the communications between the two friendly parties. Wyner's wiretap model was used to prove the existence of codes which maintain a high level of security and guarantee reliable communication between friendly parties \cite{Wyner75, Csiszar78}. Practical codes were developed by Wei \cite{Wei91}, and later examples include \cite{McLaughlin07} and \cite{Bloch06}.

Secrecy capacity $C_s$ is a fundamental limit on the rate of secure transmission \cite{Wyner75,Csiszar78}. When the main channel and the wiretap channel are modeled as BSCs, the secrecy capacity is given by $C_s = C_m - C_w$ where $C_m$ and $C_w$ are the capacities of the main channel and the wiretap channel, respectively. Strictly positive secrecy capacity implies an advantage in the quality of the main channel over that of the wiretap channel, which implies the crossover probability in the wiretap channel $p_w$ exceeds that of the main channel $p_m$ such that $0 \leq p_m < p_w \leq 0.5$. This can be the case in scenarios where distance between parties is a factor in channel quality, such as a zoned security application where friendly parties communicate inside a building and the eavesdropper monitors communications from outside of the building \cite{Harrison08}.

In applications which exhibit strictly positive secrecy capacity, tandem channel and cryptographic codes can potentially provide enhanced security. It was shown in \cite{Harrison08} that an eavesdropper can be forced to increase the number of computations needed to compromise a cryptographic system if the eavesdropper's data vector following the wiretap channel and the channel decoder still maintains a small percentage of errors. Assuming the quality of the main channel is such that the channel decoder is able to correct all of the channel errors, reliable communications can be maintained. In this paper we enhance the results of \cite{Harrison08} by showing that mutual information is correlated to the eavesdropper's ability to recover the secret key of a cryptographic system. Iterative and noniterative fast correlation techniques from \cite{Meier89} developed by Meier and Staffelbach are analyzed in an information-theoretic sense to show that \emph{cracking} the system for an eavesdropper becomes less feasible when errors from the physical layer remain uncorrected due to appropriately chosen channel codes resulting in a decrease of mutual information.

The rest of the paper is organized as follows. Section \ref{sec:algorithms} gives the background of the cryptographic attacks from \cite{Meier89} which are later analyzed to show the correlation between mutual information and system susceptibility. Section \ref{sec:information} then provides a discussion on mutual information and EXIT charts in terms of the attacks, and presents simulation results of attacks on the tandem channel and cryptographic coding scenario. Conclusions are included in section \ref{sec:conclusion}.

\section{Fast Correlation Attacks}\label{sec:algorithms}
\begin{figure*}
  \begin{center}
  \includegraphics[width=\bigPic]{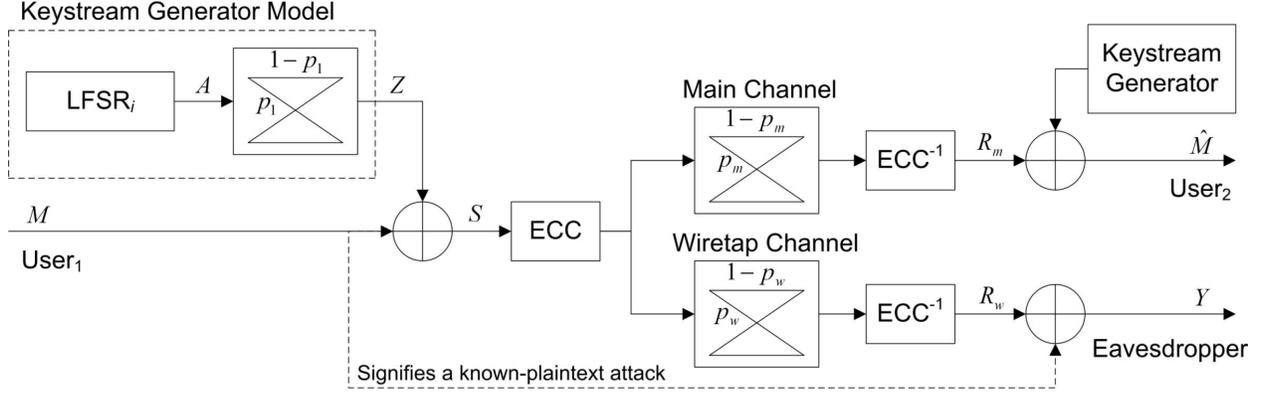}\\
  \end{center}
  \caption{Overview of wiretap channel model with channel coding and cryptography in a known-plaintext attack scenario. The keystream generator is modeled as a single LFSR preceding a BSC.}\label{fig:bigSystem}
\end{figure*}

LFSR-based keystream generators are used in many cryptographic systems, and have well-known attacks \cite{Meier89}, \cite{Siegenthaler85}, \cite{ChepyzhovS91}, \cite{JohanssonJ02}. We make use of these established algorithms in order to show how channel coding can enhance security in a wiretap setting. The encryption technique, as portrayed in Fig. \ref{fig:bigSystem}, requires a keystream generator with multiple LFSRs which need not be the same length \cite{Siegenthaler85}. Each LFSR output sequence is combined with the others using some function $f$ to form the keystream generator output $Z$. The binary message $M$ is encrypted by $S=M+Z$ where all operations are in GF(2). The sequence $S$ is encoded and transmitted. A receiver decodes the received message to obtain $R_m$, and then decrypts using an identical keystream generator which provides a perfect copy of $Z$ and calculates $\hat{M}=R_m+Z$. As long as the channel code corrects all errors due to the channel, then $\hat{M}=M$. The secret key is comprised of the initial state of each LFSR in the keystream generator.

The main assumption of fast correlation attacks on this system is that the encryption technique can be modeled such that a single LFSR output sequence $A$ is correlated with $Z$. Thus the keystream generator is portrayed in Fig. \ref{fig:bigSystem} as a single LFSR, say the $i$th one, followed by a BSC with $\Pr{(a_j\neq z_j)} = p_1$ for $j=0,1,\ldots,N-1$, where $a_j$ is the $j$th bit in the length-$N$ data sequence $A$. Fast correlation attacks on this system attempt to retrieve the initial contents of each LFSR in the keystream generator individually, and are built around the existence of checks (checksums) in $A$ which are then applied to the corresponding bits of $Z$. Checks are derived from the structure of the LFSR under attack which can be represented by a connection polynomial $g(X)=g_0+g_1X+g_2X^2+\cdots +g_kX^k$, where $k$ is the order of the polynomial and $g_j \in \{0,1\}$ for $j=0,1,\ldots,k$. All connection polynomials in the keystream generator are assumed to be primitive yielding maximal-length output sequences before repeating \cite{MoonArches}. Define $t$ as one less than the number of nonzero coefficients in $g(X)$, and denote the indices of the nonzero coefficients as $j_0,j_1,\ldots,j_t$. Then the expression $a_{j+j_0}+a_{j+j_1}+\cdots+a_{j+j_t}=0$ forms a check, and thus
\begin{equation}\label{eq:check}
  \begin{array}{ll}
    a_{j+j_u} = & a_{j+j_0} + a_{j+j_1} + \cdots + a_{j+j_{u-1}} + \\ & a_{j+j_{u+1}} + \cdots + a_{j+j_t}.
  \end{array}
\end{equation}
Almost every bit in $A$ takes part in $t+1$ checks of this kind. Additional checks are formed by squaring check expressions in GF(2) \cite{Meier89}.

Define $w$ as the total number of checks involving the bit $a_j$, and enumerate these checks from one to $w$. We now apply check expressions to bits in $Z$. Let the $v$th check be $z_j=z_{v_1} + z_{v_2} + \cdots + z_{v_t}$. Clearly $v\leq w$ because $z_j$ is in the expression. Define $b_v = \sum_{l=1}^tz_{v_l}$ and $L_v = z_j + b_v$. Then if the $v$th check holds in $Z$, $L_v = 0$. Further define $s = \Pr{(b_v=\sum_{l=1}^ta_{v_l})}$ which is the probability of an even number of bit flips in the bits $z_{v_1},z_{v_2},\ldots,z_{v_t}$. It can be shown that $s = s(t-1)$ in the recursive calculation
\begin{equation}\label{eq:sSimple}
    s(l) = (1-p_1)s(l-1) + p_1(1-s(l-1))
\end{equation}
where $s(0)=1-p_1$. Now suppose that exactly $h$ of the first $w$ checks hold in $Z$. Without loss of generality, \hfill let \hfill checks \hfill enumerated \hfill one \hfill through \hfill $h$ \hfill hold. \hfill Then \hfill we \hfill can \hfill define \\ $p^*_j = \Pr{(z_j=a_j|L_1=\cdots=L_h=0,L_{h+1}=\cdots=L_w=1)}$ which is calculated by the expression
\begin{equation}\label{eq:pstar}
  p^*_j =\frac{(1-p_1)s^{h}(1-s)^{w-h}}{(1-p_1)s^{h}(1-s)^{w-h}+p_1(1-s)^{h}s^{w-h}}.
\end{equation}
Performing the same calculation for all $N$ bits in $Z$, we form the vector $P^*=(p^*_0,p^*_1,\ldots,p^*_{N-1})$ \cite{Meier89}.

Two fast correlation known-plaintext attacks originally presented in \cite{Meier89} are now briefly summarized. Each attack assumes $N$ bits of the message $M$ are available to the eavesdropper. Furthermore, it is assumed that the eavesdropper has access to $S=M+Z$, and thus $N$ bits of $Z$ are readily available.

\subsection{Attack 1}
The first attack from \cite{Meier89} is noniterative and maintains as its motivation that those bits of $Z$ which are included in the greatest number of correct checks are more likely to be equal to their corresponding bits in $A$. Since each bit in $A$ is a linear combination of the initial state of the LFSR, it is possible to solve for the secret key using $k$ bits with linearly independent bit combinations.
An attacker selects the $k$ most reliable bits from $Z$ (i.e. the $k$ bits with the highest corresponding values in $P^*$)  that form a linearly independent system of equations. Of course if one or more of these $k$ bits are in error, then the system of equations will not return the secret key as its solution. The correctness of the solution can be determined using a threshold comparison on a correlation metric \cite{Siegenthaler85}. We consider a worst case by assuming the attacker is always able to determine whether the key obtained is correct or incorrect. If the key is incorrect, then the values of the $k$ bits are toggled trying alternate patterns with Hamming distance $1,2,\ldots,k$ until the correct key is obtained.

\subsection{Attack 2}
The second attack given in \cite{Meier89} forms an iterative update between $s$ and $P^*$, and employs two nested levels of iteration. In a particular \emph{round} of the attack, the algorithm will perform multiple \emph{iterations}. The calculation in (\ref{eq:sSimple}) is modified such that $s$ takes on a different value for each bit-check combination. A new matrix $S$ is constructed to store these values. Consider the check in (\ref{eq:check}) and call it the $v$th check. Let $(q_0,q_1,\ldots,q_{t-1})= (p^*_{j+j_0},p^*_{j+j_1},\ldots,p^*_{j+j_{u-1}},p^*_{j+j_{u+1}},\ldots,p^*_{j+j_t})$, respectively. Then the value in $S$ corresponding to the $u$th bit of the $v$th check is $S_{u,v}(t-1)$ and is calculated recursively as
\begin{equation}\label{eq:s}
  S_{u,v}(l) = q_lS_{u,v}(l-1) + (1-q_l)(1-S_{u,v}(l-1))
\end{equation}
where $S_{u,v}(0) = q_0$.
Prior to iteration $p^*_{thr}$ and $N_{thr}$ are calculated to act as decision thresholds. The calculations are based on an optimization of expected correction in the first iteration of the first round.

Each iteration computes $S$ and $P^*$ using (\ref{eq:s}) and (\ref{eq:pstar}) respectively, although (\ref{eq:pstar}) must be altered to incorporate individual values from $S$. The first calculation of $S$ assumes $p^*_j=1-p_1$ for $j=0,1,\ldots,N-1$. An attacker can estimate the channel parameter $p_1$ by counting correct checks in $Z$. If after an iteration there are greater than $N_{thr}$ elements of $P^*$ such that $p^*_j<p_{thr}$, then the round is terminated. A round consists of a maximum of $\alpha$ iterations. At the end of the round all bits $z_j$ such that $p^*_j<p_{thr}$, are flipped. All $P^*$ values are then reset to $p_1$. The attack proceeds in this fashion until it either stagnates, or converges to the correct solution. Many similarities are present between this attack and Gallager's LDPC decoding message-passing algorithm \cite{Gallager63}.

\section{Mutual Information at the Eavesdropper}\label{sec:information} With the attacks to be analyzed now well-defined, we consider Fig. \ref{fig:bigSystem} in its full context. Note that an eavesdropper has access to the data sequence $Y$ by applying the known-plaintext to the received vector $R_w$ which has already been corrected for channel errors as much as possible. We assume a strictly positive secrecy capacity and choose a channel code which allows the friendly party to correct all errors due to the main channel, but which leaves some percentage of errors due to the wiretap channel for the eavesdropper. Therefore $Y$ is a noisy version of $Z$ which can be modeled with a BSC where the probability of a bit flip from $Z$ to $Y$ is denoted $p_2$. Recall that $A$ and $Z$ are also separated by a BSC with $\Pr{(a_j\neq z_j)}=p_1$. A further simplification can occur in the modeling of the relationship between $A$ and $Y$ by combining the two cascaded BSCs into a single BSC with $\Pr{(a_j\neq y_j)}$ given as $ p' = p_1(1-p_2)+p_2(1-p_1)=p_1+p_2-2p_1p_2.$

In order to analyze per-letter average mutual information between data sequences, let the bits of each sequence be modeled as realizations of an underlying random variable. Then probability mass functions can be estimated using the available data from each sequence \cite{EXIT04}. The single BSC model with parameter $p'$ provides a convenient mechanism for analysis of information-theoretic security. Under this system, mutual information between the random variables $A$ and $Y$ is
\begin{equation}\label{eq:Ibound}
  \begin{array}{ll}
    I(A;Y) & = H(Y)-H(Y|A) \\
     & = H(Y)-H(p') \\
     & \leq 1-H(p')
  \end{array}
\end{equation}
where $H(p)$ is the binary entropy function. $H(p)$ takes its maximum value of one when $p=0.5$ \cite{Cover}. Thus as $p'\rightarrow 0.5$, then $I(A;Y) \rightarrow 0$ which effectively reduces attack 1 to a brute-force attack.

An estimate for the expected number of trials needed for attack 1 to succeed was originally derived in \cite{Meier89}, and refined in \cite{Harrison08}. Suppose attack 1 is applied to the single BSC model. The attack chooses the $k$ bits with the highest $P^*$ values which also form a linearly independent system of equations. If exactly $r$ of the $k$ bits have been flipped by the BSC, then the maximum number of trials required to cycle through all possible bit patterns up to and including $r$ errors is given by
\begin{equation}\label{eq:attackAmetric}
    A(k,r)=\sum_{i=0}^{r}{k \choose i} \leq 2^{H(r/k)k}.
\end{equation}
In practice $r$ is not known, but it can be estimated. Let $w'$ be the average number of checks relevant to any one bit, and $h'$ be the maximum integer such that $k$ bits exist which are expected to satisfy at least $h'$ checks. Then $\bar{r}$ is equal to
\begin{displaymath}
    k-\frac{k\sum_{i=h'}^{w'}{w'\choose i}(1-p')s^i(1-s)^{w'-i}}{\sum_{i=h'}^{w'}{w'\choose i}((1-p')s^i(1-s)^{w'-i}+ p'(1-s)^is^{w'-i})}
\end{displaymath}
where $s$ is calculated using $p'$ in (\ref{eq:sSimple}). Thus $\bar{r}$ of the $k$ chosen bits are expected to be in error. An estimate on the order of the number of trials required is then given as $2^{H(\bar{r}/k)k}$. This estimate is nearer to the true value when $p'$ is close to zero because when $p'\approx 0.5$ then $H(\bar{r}/k)\approx 1$ and $2^{H(\bar{r}/k)k} \approx 2^k$, but the true average in this case is only $2^{k-1}$.

\begin{figure}
  \begin{center}
  \includegraphics[width=\pic]{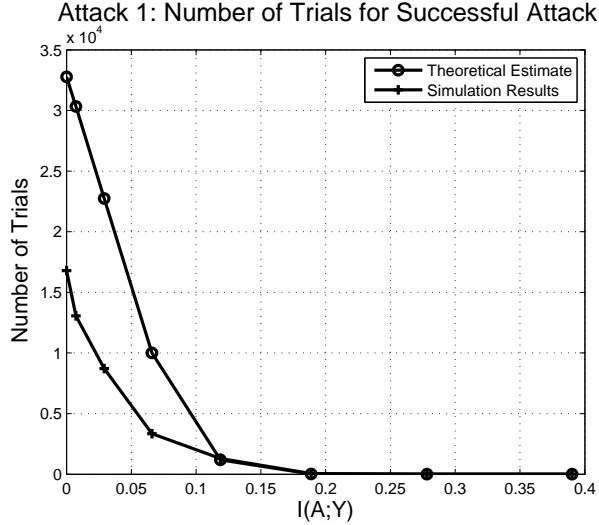}\\
  \end{center}
  \caption{Number of trials required for a successful attack versus $I(A;Y)$ using attack 1 where the order of $g(X)$ is $k = 15$, $t=4$, and $N = 1500$. Note that $t$ is small relative to $k$ for ease in simulation, but these trends extend to larger $t$.}\label{fig:MIattackA}
\end{figure}

To calculate the mutual information for attack 1, we note that since the $i$th LFSR is governed by a primitive connection polynomial, the internal state of the LFSR will take on all nonzero bit combinations. By definition the least significant bit after the $j$th shift of the LFSR is $a_j$. Thus $A$ is a maximal-length sequence which repeats after $2^k-1$ bits. If $A$ is modeled as a random variable, then $\Pr{(A=0)} = \frac{2^{k-1}-1}{2^k-1}$ and $\Pr{(A=1)} = \frac{2^{k-1}}{2^k-1}$ \cite{Welsh}. For $k$ sufficiently large, both of these probabilities are approximately 0.5. Using this density on $A$, and the crossover probability $p'$ in the aggregate BSC, $I(A;Y)$ may be calculated using the equality in (\ref{eq:Ibound}). We find that as $p'\rightarrow 0.5$ (and thus $I(A;Y) \rightarrow 0$) that the number of required iterations to recover the secret key by means of attack 1 also increases. Fig. \ref{fig:MIattackA} gives the simulated number of trials required to crack the system as a function of $I(A;Y)$ for a specific example along with the estimate from (\ref{eq:attackAmetric}).

In the case of attack 1, $I(A;Y)$ is constant throughout the attack. The information about $A$ imbedded in $Y$ is extracted and combined with knowledge of the structure of $A$ to find the secret key. However, in attack 2 the values of bits in $Y$ are modified at the end of each round, thus altering the density on $Y$ as the attack progresses. Define $Y^{[l]}$ as the $Y$ sequence after the bit flipping in round $l$ of attack 2. If the attack takes $T$ rounds for $Y^{[l]}$ to either stagnate or converge to $A$, then an information-theoretic analysis of attack 2 requires knowledge of $I(A;Y^{[l]})$ for $l=0,1,\ldots,T-1$. We expect $I(A;Y^{[T-1]})$ to equal one in a successful attack, and to be less than one in the case of a failure. A tool for viewing the expected progress of $I(A;Y^{[l]})$ is the extrinsic information transfer (EXIT) chart. EXIT charts were designed to give a graphical understanding of the decoding process of turbo codes and have been used to provide insight on the convergence of LDPC decoding as well \cite{MoonArches}. We implement EXIT analysis on attack 2 in order to characterize the expected value of $I(A;Y^{[l]})$ for $l=0,1,\ldots,T-1$, and thus display the average results of the attack under any specific channel conditions.

The mutual information between $A$ and $Y^{[l]}$ is
\begin{displaymath}
    I(A;Y^{[l]})=\sum_{a,y}p(A=a,Y^{[l]}=y)\log_2{\frac{p(A=a,Y^{[l]}=y)}{p(A=a)p(Y^{[l]}=y)}}
\end{displaymath}
where $a,y\in\{0,1\}$. Therefore, in order to calculate $I(A;Y^{[l]})$, we must estimate the probability mass function of $Y^{[l]}$, as well as the joint mass function of $A$ and $Y^{[l]}$. Since the channel we are considering is symmetric, all of this can be done through simulation by counting bits which are still in error at the end of each round and dividing by the total number of bits. Realize that the increase in information during round $l$ is $I(A;Y^{[l]})-I(A;Y^{[l-1]})$; therefore, we assign the intrinsic and extrinsic information for round $l$ as $I(A;Y^{[l-1]})$ and $I(A;Y^{[l]})$, respectively. The EXIT chart portrays the expected increase in information by plotting $I(A;Y^{[l-1]})$ versus $I(A;Y^{[l]})$ for curve one, while the second curve in the EXIT chart is $I(A;Y^{[l]})$ versus $I(A;Y^{[l-1]})$. Thus the progress of the decoder is shown by reflecting back and forth between curves. If $I(A;Y^{[l]})$ goes to one, then the attack converges on the correct sequence; therefore, there must exist a gap between the two curves if a successful attack is to be expected.

In order to show average tendencies in the mutual information during attack 2, we construct EXIT charts using a binning technique. The mutual information $I(A;Y^{[l]})$ is calculated for every round in a large number of attacks. Then the expected increase in information is obtained for each section of the chart by subdividing the x-axis into $D$ equal segments or bins. The data are sorted according to intrinsic information, and then the extrinsic information is averaged in each bin. The center of each of the $D$ segments is used as the intrinsic information for the corresponding bin when forming the chart.

\begin{figure}
  \begin{center}
  \includegraphics[width=\pic]{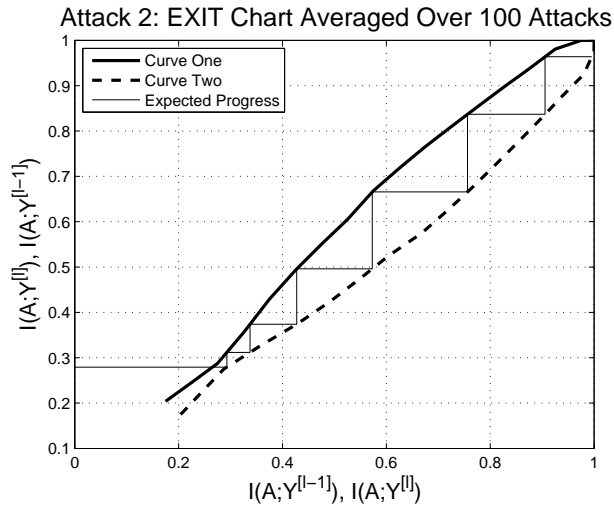}\\
  \end{center}
  \caption{EXIT chart with $D=20$ formed by averaging the results of 100 simulations of attack 2 with $k=31$, $t=6$, $\alpha=5$, $N=3100$, and $p_1=p'=0.2$.}\label{fig:EXIT2}
\end{figure}

Results using this method for a particular set of system parameters averaged over 100 attacks are shown in Fig. \ref{fig:EXIT2}. For this example, we assume that the eavesdropper corrects all errors due to the physical layer yielding $p_2=0$. We observe that the EXIT chart predicts an overall tendency for the attack to succeed due to the gap between curves. We also note that the EXIT curves do not extend to zero. Generating these curves was implemented by actually simulating attacks on the system. Although some rounds did yield a negative correction, none resulted in zero extrinsic information; therefore, no rounds exhibited zero intrinsic information either, leaving bins around zero empty. Finally we observe that the gap between EXIT curves is narrower for lower intrinsic information regimes. This fact defends the technique used in \cite{Meier89} and \cite{Harrison08} in defining the \emph{correction capability} of attack 2 using only the expected results in the first round of an attack. If the first round provides good correction, then the chart indicates that convergence to $A$ will proceed quickly. When the first round has mediocre or poor correction, the algorithm must proceed through the pinched region of the gap resulting in slower convergence.

\begin{figure}
  \begin{center}
  \includegraphics[width=\pic]{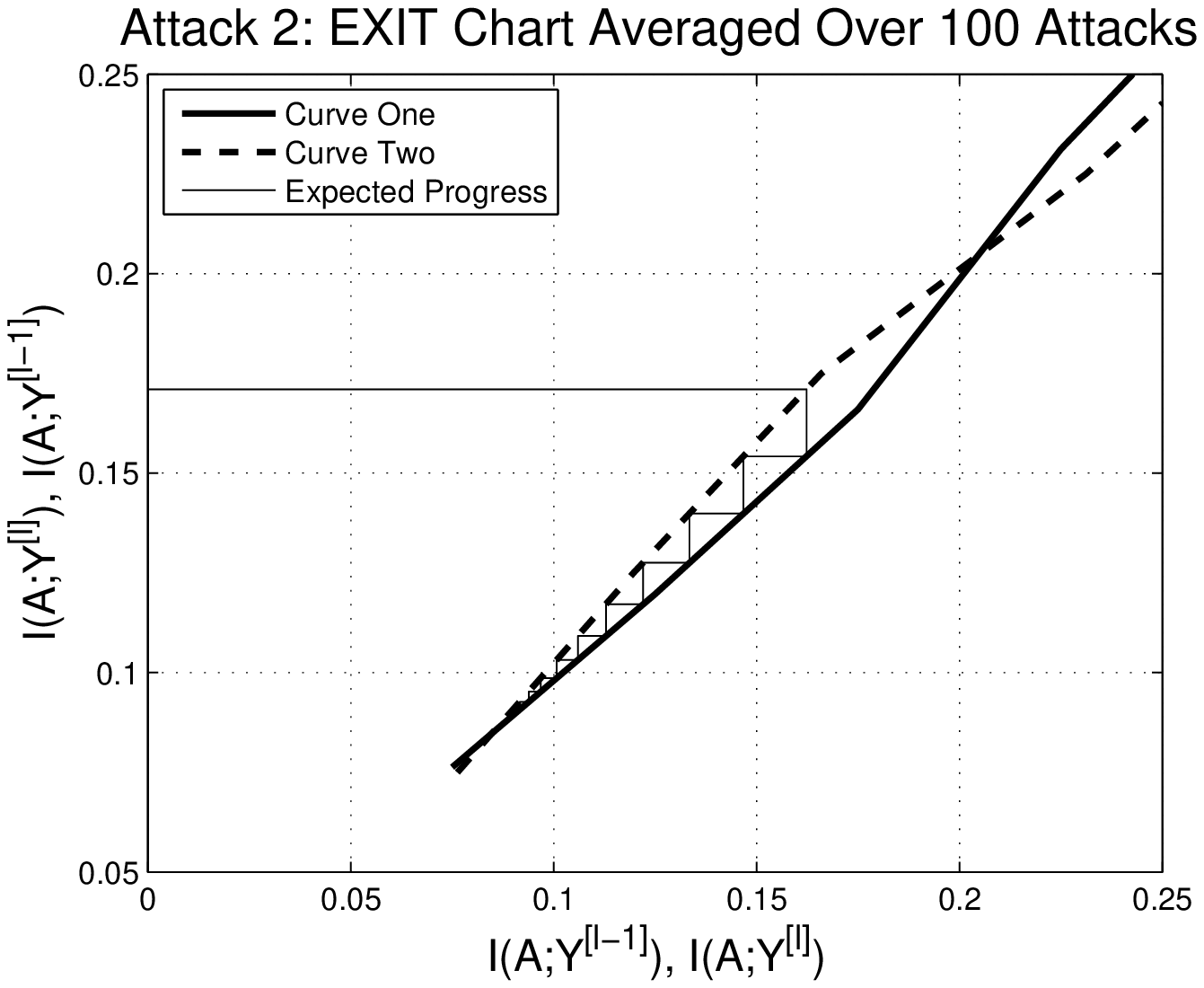}\\
  \end{center}
  \caption{EXIT chart with $D=20$ formed by averaging the results of 100 simulations of attack 2 with $k=31$, $t=6$, $\alpha=5$, $N=3100$, $p_1=0.2$, and $p_2=0.1$ yielding $p'=0.26$.}\label{fig:EXIT3}
\end{figure}

Another EXIT chart for a similar setup as that in Fig. \ref{fig:EXIT2} is shown in Fig. \ref{fig:EXIT3}. The only difference in the two scenarios is the nonzero value of 0.1 for $p_2$ in the attacks depicted by Fig. \ref{fig:EXIT3}. This figure shows that attack 2 is likely to fail due to the crossover in the EXIT chart when a channel code is assumed which delivers a 10\% error rate to the eavesdropper. Again this behavior can be predicted from the average correction in the first round. Fig. \ref{fig:EXIT3} portrays more errors on average following the first round than there were prior to launching the attack. In this scenario, the expected progress of an attack converges on the crossover point in the EXIT chart rather than converging to one as in Fig. \ref{fig:EXIT2}. Thus an attack that would be otherwise successful can be expected to fail if a channel code can be used to ensure enough errors in $R_w$ to create a crossing point in the EXIT chart.

\section{Conclusion}\label{sec:conclusion} In conclusion, we see that mutual information is a meaningful metric in developing the concept of information-theoretic security. It is verified using mutual information and EXIT charts that fast correlation attacks can be made more difficult or impossible when channel errors at the physical layer are considered. Channel errors can be brought to bear against an eavesdropper through implementing security-enhancing channel codes. Noniterative attacks which effectively shrink the average search area to find the key can expect an increase in necessary computations for successful decoding. EXIT charts provide an excellent analysis tool in determining the increase in security due to channel codes when iterative cryptographic attacks are employed, signifying an average failure to decode when the curves in the EXIT chart cross. Finally EXIT analysis may also provide more insight into optimizing attacks on the LFSR-based encryption scheme presented.


\bibliographystyle{unsrt}
\bibliography{LFSRreferences}

\end{document}